\documentclass[12pt]{article}

\usepackage{pdfpages}
\usepackage{./scicite}

\usepackage{times}
\usepackage{graphicx}% Include figure files
\usepackage{placeins}
\usepackage[labelfont=bf]{caption}

\newenvironment{sciabstract}{%
	\begin{quote} \bf}
	{\end{quote}}

\newcounter{lastnote}

\title{Nanofluidic Rocking Brownian Motors}% Force line breaks with \\  % or ratchet?

\author
{Michael J. Skaug,$^{1,+}$ Christian Schwemmer,$^{1,+}$ Stefan Fringes,$^{1,+}$  
	\\
	 Colin D. Rawlings,$^{1}$ Armin W. Knoll$^{1\ast}$\\
	\\
	\normalsize{$^{1}$IBM Research-Zurich,}\\
	\normalsize{S\"aumerstrasse 4, 8803 R\"uschlikon, Switzerland}\\
	\normalsize{$^+$Equally contributing authors}\\
	\normalsize{$^\ast$To whom correspondence should be addressed; E-mail:  ark@zurich.ibm.com}	
}

\date{}

\begin{document} 
	
	
	%\small
	%\twocolumn
	%\baselineskip12pt
	
	\baselineskip15pt
	
	
	\maketitle

	
\begin{sciabstract}
Control and transport of nanoscale objects in fluids is challenging because of the unfavorable scaling of most interaction mechanisms to small length scales. We design energy landscapes for nanoparticles by accurately shaping the geometry of a nanofluidic slit and exploiting the electrostatic interaction between like charged particles and walls. Directed transport is performed by combining asymmetric potentials with an oscillating electric field to achieve a rocking Brownian motor. Using 60\,nm diameter gold spheres, we investigate the physics of the motors with high spatio-temporal resolution, enabling a parameter-free comparison with theory. We fabricated a sorting device that separates 60- and 100-nanometer particles in opposing directions within seconds. Modeling suggests that the device separates particles with a radial difference of 1 nanometer.
\end{sciabstract}

%One Sentence Summary: A 3D-shaped nanofluidic confinement and an oscillating electric field enable size selective nanoparticle transport.

% % % Intro % % % % % % % % % % % % % % % % % % % % % % % % %
%
%
%Main Text: 
Lab-on-chip devices that can size-selectively transport and collect nanoscale particles are expected to find applications in materials and environmental science [e.g., size analysis, filtration, and monodisperse production \cite{salafi2017advancements,Marago2013natnan}] as well as in point-of-care diagnostics and biochemistry [e.g. molecular separation and pre-concentration \cite{fu2007patterned,wunsch2016nanoscale,napoli2010nanofluidic,salafi2017advancements}]. For example, directed transport may overcome fundamental limits in the detection of dilute species in fluidics \cite{Squires2008Making} by actively transporting and accumulating them at the sensor area. Inspired by molecular motors in biology,  Magnasco \cite{Magnasco1993} and Prost \textit{et al.} \cite{Prost1994} proposed that such particle transport could be achieved with artificial Brownian motors (BMs) based on an asymmetric energy landscape and non-equilibrium fluctuations. Previous experiments \cite{Faucheux1995,lee2005observation,Rosselet1994Nature,Marquet2002,bogunovic2012particle,Bader1999} focussed primarily on "flashing ratchet"-type BMs that exploit a periodically generated asymmetric trapping potential and isotropic diffusion to transport micrometer-scale particles. The required potentials were obtained using optical \cite{Faucheux1995,lee2005observation} or dielectrophoretic \cite{Rosselet1994Nature,Marquet2002,bogunovic2012particle} forces, which scale with the particle volume and therefore are not efficient at the nanoscale. Accordingly, direct charge-charge interactions were required to transport DNA molecules using intercalated electrodes \cite{Bader1999}. 

Such 'flashing' BMs were also explored for particle sorting by exploiting the dependence of diffusion on particle size. However, similar to the case of continuous-flow devices \cite{Huang14052004,wunsch2016nanoscale}, it is expected that using external forcing instead of diffusion will result in greater separation precision. Rocking BMs \cite{Magnasco1993,Astumian1997,Marquet2002} use a zero-mean external force and a static potential landscape to generate directed particle motion. Their transport exhibits a strong nonlinear dependence on particle diffusivity, which is promising for nanoparticle separation  \cite{Bartussek1994}. However, for nano-scale particles, creating a sufficiently strong static energy landscape remains a challenge. 

Electrostatic trapping \cite{Krishnan10nature} addresses this challenge by confining like-charged particles between uniformly charged surfaces. A geometrical recess in one of the surfaces lowers the particle-surface interaction energy locally and thus defines a lateral trapping potential. Confinement energies of several multiples of $k_B T$  (where $k_B$ is the Boltzmann constant and $T$ is absolute temperature) were achieved, resulting in the stable trapping of various types of charged nanoparticles, such as 10-100 nm diameter Au nanospheres \cite{Kim14natcom,Krishnan10nature,Gerspach15}, Au nanorods\cite{Celebrano12nanlet}, 50 nm diameter vesicles \cite{Krishnan10nature}, and even 10- to 60-base DNA oligonucleotides and 10 kDa proteins \cite{RuggeriZoselMutterEtAl2017}. External forcing by optical and electrical fields was also explored as a means of driving the particles from one stable minimum to another \cite{myers2015information}. 

We expand the concept of geometry-induced electrostatic trapping to create complex two-dimensional (2D) energy landscapes for nanoparticles by replacing the simple recess geometries with lithographically patterned 3D topographies. The patterns were fabricated using thermal scanning probe lithography (t-SPL) \cite{pires2010nanoscale}, which has a depth accuracy in the nanometer range \cite{knoll2017control}. We implement nanofluidic rocking BMs for our model particle system of 60-nm Au nanospheres. 
We created energy landscapes of up to $\approx 10\,k_B T$ in scale at a lateral resolution of less than 100 nm. 
Using high spatio-temporal resolution optical microscopy we determined all relevant physical system parameters in-situ, such as the $\approx 10\,$nm spatially resolved particle interaction potential and the millisecond-resolved particle motion. We find excellent agreement with theory and demonstrate a sorting device.

% % % Experimental % % % % % % % % % % % % % % % % % % % % % % % % %
%
%

\begin{figure} [ht]
	\centering
	\includegraphics[width=15cm]{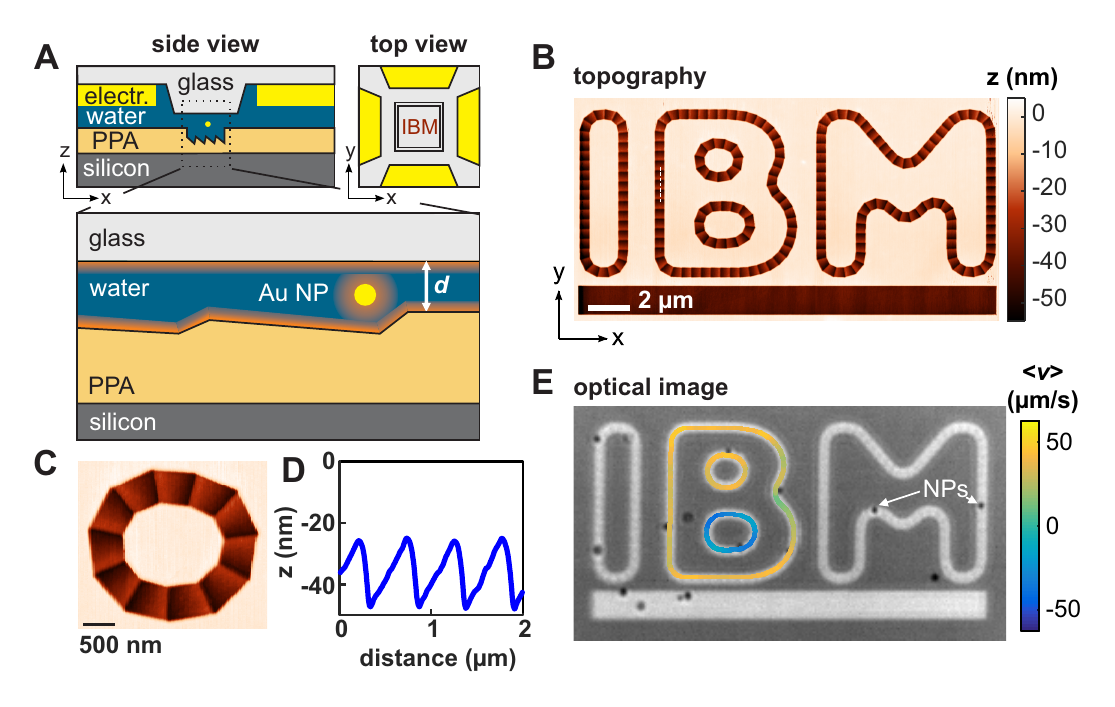}%
	\caption{\label{schematic} \textbf{Nanofluidic Brownian motor setup and ratchet topography defined by thermal scanning probe lithography (t-SPL).} \textbf{(A)} Schematic cross section and top view of the nanofluidic slit. Top left: A pillar (heigt 30 to 50 $\mu$m) etched into the cover glass provides close proximity, optical access, and space for electrodes. Top right: The top view (not to scale) depicts the electrodes (yellow, spaced by $\approx 1.2\,$mm) and the pillar in the center (width $\approx 100\,\mu$m). Two voltage signals $U_{x,y}(t)$, zero-mean-square-shaped with a phase shift of 90$^o$, are applied in $x$ and $y$ directions to the electrodes, thus creating a rotating electric field that drives the motor. Bottom: The ratchet topography in polyphthalaldehyde and the $60\,$nm gold nanoparticle, drawn to scale. The particle experiences a ratchet-shaped energy landscape due to the ion-cloud interactions (orange) between like-charged surfaces. The gap distance was $d = 150\pm1\,$nm. \textbf{(B)} Topography image of the patterned geometry. \textbf{(C)} Close-up of the small circular ratchet shape of (B). 
		\textbf{(D)} Cross section of the track's sawtooth geometry measured along the white dashed line in (B). \textbf{(E)} Optical image of particles trapped in the ratchet. The color overlay indicates the average speed measured along the ratchet path in the counterclockwise direction at an applied rotating voltage of 3\,V\,$@$\,30\,Hz.}
\end{figure}

Gold nanoparticles in electrolyte 
% (Debye length $\kappa^{-1} = 15\pm3\,$nm) 
were confined in a nanofluidic slit of controllable gap distance $d$ (Fig. \ref{schematic}\,A and fig.  S1) \cite{MM} using the nanofluidic confinement apparatus described in \cite{fringes2018diff}. We used closed-loop t-SPL \cite{knoll2017control} to pattern the thermally sensitive polymer polyphthalaldehyde with a sawtooth profile along complex-shaped transport tracks having a depth of 30 to 50 nm and a period of $L \simeq 550$ nm (Fig. \ref{schematic},\,B to D). 
%The surfaces of the nanoparticles, the cover glass and the polymer were negatively charged \cite{fringes2016diff}. At separations of a few Debye lengths $\kappa^{-1} $, the particles are repelled from the like charged confining surfaces. 
The modulated gap height (Fig. \ref{schematic}\,A) in the slit results in a static lateral particle energy landscape, which confines the particles to the tracks (Fig. \ref{schematic}\,E) and provides the static asymmetric potential required for the rocking BM. 

\begin{figure*}[h]
	\centering
	\includegraphics[width=15cm]{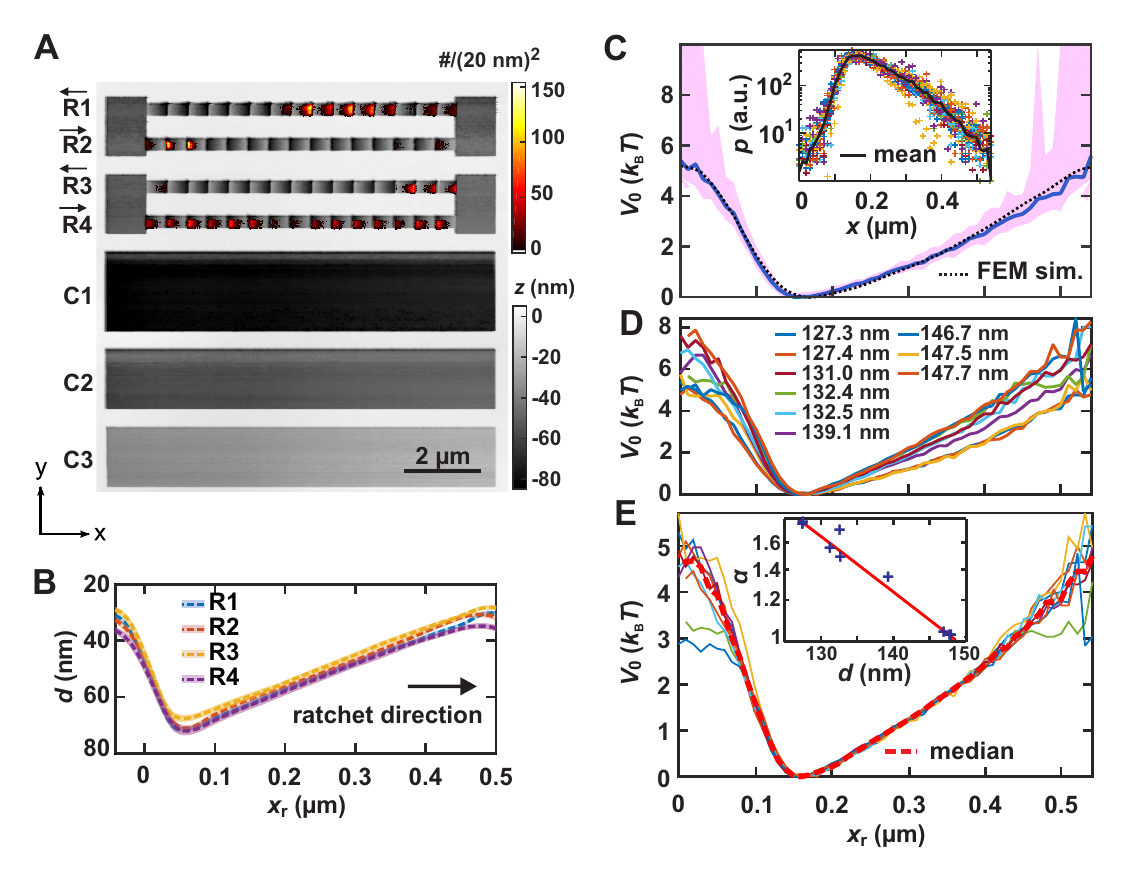}%
	\caption{\label{potentials}\textbf{Experimental determination of the average energy landscape.} \textbf{(A)} Topography image of the patterned ratchets R1 to R4 and three control fields, C1 to C3. Arrows indicate the ratchets' direction. The observed particle positions are overlaid as a heat map. \textbf{(B)} Single-cell averaged cross sections (dashed lines) of the four ratchets including standard deviations (areas, 15 cells per ratchet) along the ratchet direction $x_r$. \textbf{(C)} Average experimental 1D energy profiles $V_0(x_r)$ of the particles in the ratchet compared to a profile obtained from finite element modeling (FEM; dotted line). The blue line denotes the average interaction energy as estimated from the average occupation probability (inset, 24 profiles) using Boltzmann's principle. The shaded region marks the standard deviation. \textbf{(D)} $V_0(x_r)$ for nine gap distances of $127\,$ nm$\,< d < 148$\, nm (standard deviation of $d$, $\approx 1\,$nm). \textbf{(E)} Master curve obtained by linear scaling of the potentials by a factor of $\alpha$. The inset depicts the exponential dependence of $\alpha$ on $d$.}
\end{figure*}

\FloatBarrier

To elucidate the effect of the topography on the particle's interaction energy and to demonstrate the operation of the BMs we resorted to simple linear ratchets (Fig. \ref{potentials}\,A). 
%The BMs transport particles betweeen compartments on the left and right of the ratchets. 
We define the direction of the BMs as the direction of the shallow slope profile (Fig. \ref{potentials}\,B), indicated by the arrows in Fig. \ref{potentials}\,A. In addition three control fields C1, C2, and C3 were written with a depth of the maximum, mean, and minimum depth of the ratchets, respectively. The control fields are used for in situ measurement of the mean particle diffusivity and the mean particle speed in the non-driven and driven case, respectively. 

We recorded the position of the particles with an effective particle illumination time of $\approx 40 \mu$s and a frame rate of 1000 fps \cite{fringes2018diff} using interferometric scattering detection microscopy \cite{lindfors2004detection,fringes2018diff} (movie S1).
The overlaid heat map (2D histogram) in Fig. \ref{potentials}\,A indicates the density of all recorded particle positions.
Owing to the brief illumination time, the particle positions obtained from each frame can be used directly to obtain particle occupation histograms \cite{MM}, (fig.\,S2). 
We collapsed the observed 2D particle positions onto the $x$-axis and calculated an average 1D histogram for a single ratchet cell by cross-correlation averaging. The normalized histograms of the single cells are shown in the inset of Fig. \ref{potentials}C. Identifying the histograms with the average 1D occupation probabilities $p(x)$,  we calculate the effective 1D interaction energy $V_0(x)$ of the particles using Boltzmann's principle $p(x) \propto exp[- V_0(x)/k_B T]$ (Fig. \ref{potentials}\,C). 

In the nanofluidic confinement apparatus the gap distance is controlled and measured \cite{fringes2018diff} with nanometer accuracy, which provides a convenient handle to tune the energy scale. Figure \ref{potentials}\,D shows nine recorded energy profiles for $127\,$nm$\,< d <\,148\,$nm. Within this range the curves are self-similar and collapse on a master curve obtained by linear scaling (Fig. \ref{potentials}E). The scaling coefficient $\alpha$ decreases exponentially $\alpha \propto exp(-d/2\,l)$, where $l = 18 \pm 3\,$nm. Finite element modeling simulations\cite{MM} (figs. S3 and S4) yield similar potential shapes (Fig. \ref{potentials}\,C) and indicate that $l$ is slightly greater than the Debye length $\kappa^{-1} \approx 12\,$nm \cite{MM} (table S1).

\begin{figure}
	\centering
	\includegraphics[width=15cm]{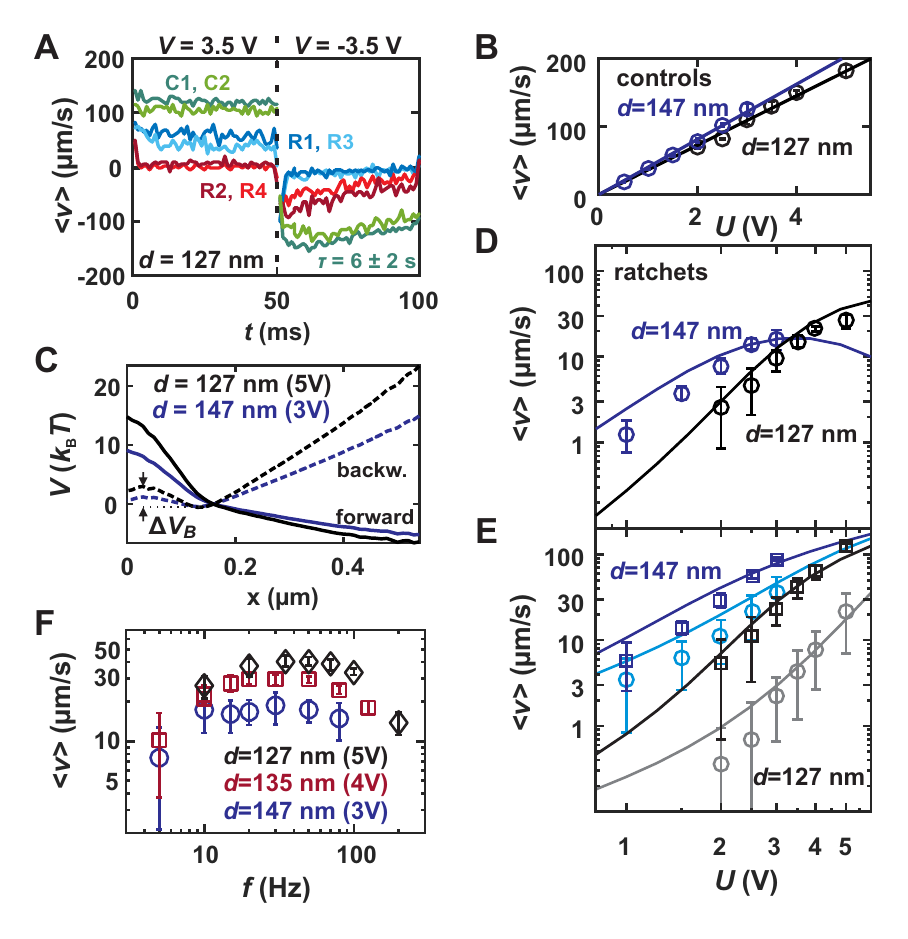}%
	\caption{\label{fig_operation}\textbf{Brownian motor operation.} \textbf{(A)} Time-resolved average velocities measured in the control fields C1 and C2 and ratchets R1-R4. \textbf{(B)} Average particle speed in the control fields as a function of the applied voltage $U(V)$ and for two gap distances $d = 147\,$nm and $127\,$nm. \textbf{(C)} Estimated forward (solid lines) and backward (dashed lines) 1D potential energy $V(x_r)$ for the two gap distances and applied voltages of 3 and 5\,V. Arrows indicate the magnitude of $\Delta V_B$. \textbf{(D)} Average measured particle drift velocity in ratchets R1 to R4. \textbf{(E)} Average particle drift velocities shown separately in the forward (blue and light blue) and backward directions (black and gray) of the ratchets. Lines in (D) and (E) indicate a fit-parameter-free comparison with theory. \textbf{(F)} Average overall drift velocity $\left<v\right>$ observed in ratchets R1 to R4 as a function of the driving frequency for three gap distances and their respective maximum voltages.  Error bars in (B), (D), (E),
	and (F) denote SD of measurements obtained for each (ratchet) field.}
\end{figure}

The patterned nanofluidic slit thus defines an asymmetric static potential landscape for the particles - one of the two required ingredients for a rocking Brownian motor. The other necessary ingredient is non-equilibrium fluctuations, which we created by applying a 10-Hz electric square-wave potential to gold electrodes (Fig.\ref{schematic}\,A and fig. S1). 
The electric field induces both an electrophoretic force on the nanoparticles and an electro-osmotic plug flow of the electrolyte, acting on the particles in opposite directions. Our experiments showed that electro-osmotic effects dominate because the particles are dragged toward the negative electrode. Fig. \ref{fig_operation}\,A depicts the particle drift velocities measured at time intervals of 1 ms,  $d=127\,$nm, and $U=3.5\,$V $@$ 10 Hz, resolving each half cycle of the electric field. 

We use the average drift velocity (fig. S5) in the control fields, $\left\langle v_c\right\rangle$, to quantify the total force as a function of the gap distance (Fig. \ref{fig_operation}\,B).
Combining Einstein's relation $D_0 = k_B T / 6 \pi \eta a$ and Stoke's equation for particle drag $F = 6 \pi \eta a \left\langle v_c\right\rangle$ (where $\eta$ is dynamic viscosity $a$ is particle radius) yields $ F/k_B T = \left\langle v_c\right\rangle/D_0$. At $U=5\,$V and $d=127\,$nm, we observe $F=39 \pm 2\, k_B T/\mu$m for the measured average diffusion coefficient $D_0 = 4.6 \pm 0.2\,\mu$m$^2$/s (SD of 18 measurements). 
Maximum voltages were restricted to 5\,V at $d=127\,$nm and 3\,V at $d=147\,$nm (movies S2 and S3) to prevent particles escaping from the reservoirs. 

In each half cycle of the electric field, the force is approximately constant, which corresponds to a linear potential being added to the static potential $V_0$: $V(x) = V_0(x) \pm x F$. Fig. \ref{fig_operation}\,C shows the tilted potentials for the most and least-confined experiments ($d = 127\,$nm and $d = 147\,$nm) and the maximally applied voltages of 5 and 3\,V, respectively. In the forward direction, all energy barriers disappear and the particles begin to drift. In the backward direction, however, finite energy barriers $\Delta V_B$ remain, thus hindering drift. Oscillating the force leads to a ``rocking'' of the asymmetric particle potential, resulting in directed transport due to the rectification of particle drift.

According to Reimann \textit{et al.}, the average drift velocity can be computed for the tilted potentials shown in Fig. \ref{fig_operation}C using the first passage time model \cite{Reimann2002a}:
\begin{eqnarray}
\label{eqxD}
\left<v\right> &=& \frac{1-e^{-L F / k_B T}}{\int_{0}^{L} L^{-1}\,I_\pm (x) \, dx} \label{vdrift} \\
I_\pm (x) &=& \frac{1}{D_0} \int_{0}^{L} \exp\left\{ \pm [V_0(x) - V_0(x \mp z)]/k_B T \right\} dz
\end{eqnarray}

All relevant, non-trivial quantities $V_0(x), D_0,$ and $F$ in this theoretical description were measured in-situ, and a parameter-free comparison with theory is possible. Equation \ref{eqxD} was evaluated numerically and the total drift was calculated as one-half of the drift difference in forward and backward directions (lines in Fig. \ref{fig_operation}\,D and E). The theoretical description assumes slow rocking; that is, the actuation time of each half-cycle is long relative to the particle-drift time across a single ratchet element. Our experiments show excellent agreement with the model at 10 Hz (fig. S6). Moreover, we found a frequency dependence of the ratchet performance (Fig. \ref{fig_operation}F). Best performance is observed when the actuation time at 30 to 50 Hz is about twice the particle-drift time $L/\left<v\right> \approx 6\,$ms in the forward direction (movies S4 to S6). 

For the curved paths shown in Fig.\,\ref{schematic}C, a rotating electric field was applied to achieve efficient operation (fig. S7 and movie S7). The electric field direction can also be used to selectively power ratchets written in orthogonal directions, for example to shuttle particles between several reservoirs (fig. S8 and movie S8). 

\begin{figure*}[h]
	\centering
	\includegraphics[width=15cm]{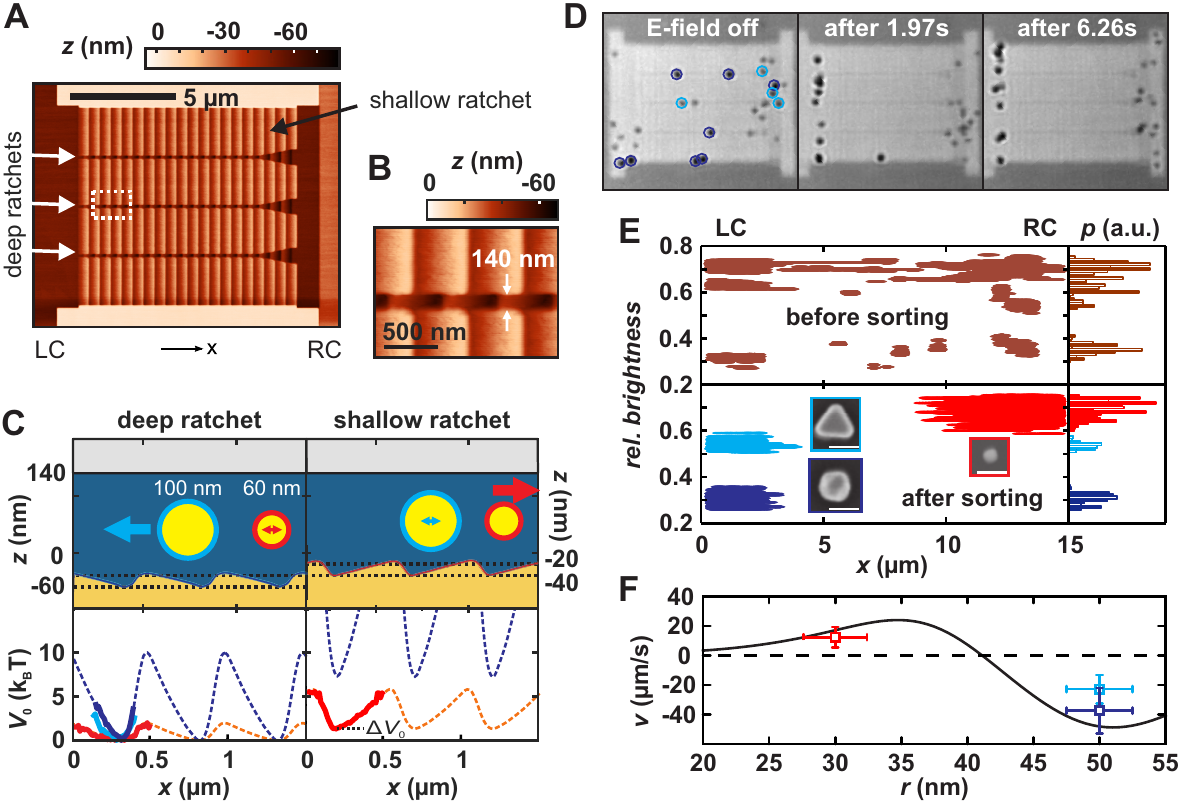}%
	\caption{\label{fig_sorting}\textbf{Sorting of 60- and 100-nm Au particles.} \textbf{(A)} Topography of the sorting device as determined by t-SPL. Three deep ratchets (white arrows) are embedded in a wide shallow ratchet for transporting the 100- and 60-nm particles into compartments LC and RC, respectively. The deep ratchets comprise wide ends for better entrance of nanoparticles from RC. \textbf{(B)} Close-up of the dotted white box (A) depicting the central deep ratchet (width 140\,nm). \textbf{(C)} Top: Schematics of the sorting device, roughly to scale with experimental conditions. The 60-nm particles (red) in the shallow ratchet and the 100-nm particles (blue) in the deep ratchet are transported in opposite directions (arrows). Bottom: The corresponding measured (solid) and modeled (dashed) static energies. The 60\,nm particles experience an energy barrier of $\Delta V_0 \approx 1.3\,k_B T$ as they move into the shallow ratchet. \textbf{(D)} Optical microscopy images before sorting, and after 2\,s and 6.3\,s of applied voltage of 4\,V\,$@$\,30\,Hz AC. Colored circles indicate the 100-nm particle types depicted in (E). \textbf{(E)} Spatial particle distribution (left) and histograms (right) of the median relative brightness of the particle trajectories measured before and after sorting. We observe a third particle population (light blue), which we attribute to platelets according to their relative abundance. Insets: SEM images of the corresponding particles (scale bars, 100 nm). \textbf{(F)} Modeled average particle drift $v$ as a function of particle radius $r$. The data points indicate the measured speed (mean $\pm$ SD) of 15 particles. The radial error corresponds to the measured SD (60 nm: 8\%; 100 nm: 5\%) of the particle size determined by SEM.\cite{MM}
	}
\end{figure*}

Finally, we exploit the strong nonlinear gap-distance dependence of the energy landscape to separate 60- and 100-nm particles into opposite compartments, LC and RC, of a sorting device (Fig.\,\ref{fig_sorting}A and B). 
The middle area between the compartments contains three deep (local $d = 185 - 210\,$nm) and narrow (width $\approx 140$\,nm) ratchets pointing to the left and embedded in a wide (8 $\mu$m) and shallow (local $d = 160 - 190\,$nm) ratchet pointing to the right.  
Our experiments show that at a gap distance of $d = 150\,$nm, the big particles are well confined to the deep ratchets (movie S9). Although the small particles experience an energy barrier of $\Delta V_0 \geq 1.3\,k_B T$ (Fig. 4C and fig. S9) to move from the deep ratchet into the shallow ratchet,
its wide area (4$\mu$m $\times$ 2\,$\mu$m vs. 3\,nm $\times$ 140\,nm) leads to a measured occupation probability of 76\% in the shallow ratchet. Combined with the measured ratchet potential of $\approx 4.5\,k_B T$  (red line in Fig. \ref{fig_sorting}C), this leads to a transport of $60$-nm particles toward RC. 

The optical images in Fig. \ref{fig_sorting}D demonstrate the efficient sorting of the particle populations in opposite directions. Initially, the particles are distributed randomly across the ratchet field. Applying a square-wave voltage of 4\,V $@$ 30 Hz, we found that the populations were fully separated into the respective compartments after $2\,$s, except for a single 100 nm particle stuck in the shallow ratchet. After 6.3\,s, this particle diffused into the deep channel and was transported into the LC. Figure \ref{fig_sorting}E depicts the median brightness of measured particle trajectories in the left and right halves of the field. The histograms at the right in Fig. \ref{fig_sorting}E indicate that next to the majority particle types of 60-nm (red) and 100-nm (blue) spheres, there is a small population of a third particle type (light blue) in the sample. Inspection by scanning electron microscopy (SEM) revealed that, apart from the spheres,  the 100-nm dispersion contained a small fraction of ``platelets'', which we assign to the intermediate particle type \cite{MM} (figs. S10 and S11). Using the median trajectory contrast for differentiation, we measured a  particle potential for this population of $\approx 6\,k_B T$ in the deep ratchet (Fig. \ref{fig_sorting}A). Note that although the brightness histograms of 60-nm spheres and 100-nm platelets almost overlap, all platelets were actively driven toward LC (movie S9).

The operation and separation potential of the device can be understood using a simple electrostatic model \cite{Behrens2001,MM} to calculate the particle energy (dashed lines in Fig. \ref{fig_sorting}C), assuming a linear scaling of the applied force with particle radius, and using equation \ref{eqxD} to calculate the particle speed as a function of radius \cite{MM}. The model yields an average drift speed, which is compared to the measured speeds in Fig.\,\ref{fig_sorting}F. With increasing particle size the deep ratchets produce an increasingly dominating bias of a negative drift speed. This is due to a stronger confinement to the deep ratchets and an increasing ratchet potential experienced by the particles in the deep ratchets (from $\approx\,2\,k_B T$ for the 60-nm particles to $\approx\,10\,k_B T$ for the 100-nm particles, Fig.\,\ref{fig_sorting}C). This leads to a zero crossing of the net drift at a particle diameter of $\approx\,82\,$nm ($r = 41$\,nm). The strong nonlinear character of the curve originates from the intrinsically nonlinear BM transport and the exponentially increasing interaction energy, resulting in a slope of $-7\,\mu m\,s^{-1} nm^{-1}$ at the zero crossing. Thus, two particle species with a radial difference of $1\,$nm would drift with $> 3\,\mu m\,s^{-1}$ in opposite directions, which is sufficient to overcome the diffusion current that drives the particles back into the device \cite{MM} (figs. S12 and S13).

\FloatBarrier

Our results show that confining nanoparticles in electrolyte between a flat and a topographically patterned surface creates an energy landscape defined by the topography. By tuning the gap distance, the potential landscape to first order is simply scaled in magnitude, providing a convenient handle to optimize the system. All relevant physical quantities necessary to model the system are accessible in situ. Agreement between theory and experiment proves the validity of the interpretation and the predictability of the system. The nonlinear character of rocked BM transport \cite{Bartussek1994} and of the electrostatic interaction leads to efficient separation \cite{MM} by our device, similar to the case of other concepts based on geometrical constrictions \cite{fu2007patterned,derenyi1998ac}. 

Moreover, our modeling \cite{MM} and the trapping results of Ruggeri et al. \cite{RuggeriZoselMutterEtAl2017} indicate that the method should scale to relatively small biomolecules. 
In contrast to flow-based separation \cite{Huang14052004}, the rocked Brownian motor implementation provides selective transport, precise separation, and accumulation of nanoparticles without a net flow of the electrolyte or sustained thermodynamic gradients. Combined with the small footprint and the low applied voltage, such devices are ideally suited for the precise analysis of small liquid volumes in lab-on-chip technology.

%\bibliography{BMpaper}

%\bibliographystyle{Science}

%\begin{scilastnote}
	%\item 
\section*{Acknowledgments}
	
	We thank U. Drechsler for assistance in fabricating of the glass pillars, U. Duerig and H. Wolf for stimulating discussions, and R. Allenspach and W. Riess for support. We thank C. Bolliger and L.-M. Pavka for proofreading the manuscript. \textbf{Funding}: Supported by European Research Council Starting Grant 307079, European Commission FP7-ICT-2011-8  no. 318804, and Swiss National Science Foundation grant 200020-144464. \textbf{Author contributions:} M.J.S. and A.W.K. jointly conceived the idea and the experimental concept. M.J.S. implemented first working devices and started theoretical analysis. C.S. and S.F. performed the sorting experiment. C.S. performed the analysis of the sorting experiment and the numerical modelling for the BMs and the sorting experiment. S.F. performed the BM experiments and partially their analysis. C.D.R. performed the finite element modeling and analysis. A.W.K. performed the t-SPL lithography, analyzed the BM function, supervised the work, and wrote the manuscript. C.S., C.D.R., S.F. and A.W.K. wrote the Supplementary Materials. \textbf{Competing interests:} The authors declare no competing interest. \textbf{Data and materials availability:} All data needed to evaluate the conclusions in the paper are present in the paper and/or the Supplementary Materials. 
	%Additionally, the figures including the experimental data are available at https://ibm.box.com/s/a71dg6vhbwe6q2ktrqaeggs1c6tx8ho1. 
	All authors are inventors on US patent application (15/469995) submitted by IBM that covers transport and separation of nanoparticles.

\section*{Supplementary materials}

www.sciencemag.org/content/359/6383/1505/suppl/DC1

Materials and Methods
Supplementary Text\\
Figs. S1 to S13\\
Table S1\\
Movies S1 to S9\\
References \textit{(32 - 37)}



\section*{Supplementary material}

# Supplementary Materials for

## Nanofluidic Rocking Brownian Motors

Michael J. Skaug, Christian Schwemmer, Stefan Fringes, Colin D. Rawlings,
Armin W. Knoll

IBM Research – Zurich, 8803 Rüschlikon, Switzerland

Correspondence to: ark@zurich.ibm.com

**This PDF file includes:**

    Materials and Methods
    Supplementary Text
    Figs. S1 to S13
    Table S1
    Captions for Movies S1 to S9

**Other Supplementary Materials for this manuscript includes the following:**

    Movies S1 to S9

**Materials and Methods**

Nanoparticles

We used citrate-stabilized 60 nm and 100 nm Au nanospheres from BBI Solutions for our experiments. According to the manufacturer, the 60 nm particles have a particle concentration of $2.6 \times 10^{12} \, \text{cm}^{-3}$ and the 100 nm particles a concentration of $5.6 \times 10^{9} \, \text{cm}^{-3}$. Their coefficient of variation for the nanoparticle diameter was inferred from scanning electron microscopy (SEM) pictures. It was 8 % for the 60 nm particles and 5 % for the 100 nm spherical particles. The 100 nm particle dispersion also contained a ≈20 % fraction of platelets, see Fig. *S10*. To achieve an optimal number of particles in our experiments, we diluted the 60 nm dispersion with ultrapure water (Millipore, 18 MΩcm) and concentrated the 100 nm dispersion with a centrifuge immediately prior to the experiments. A summary of all dispersions that were used for our experiments can be found in Table *S1*.

To further characterize the nanoparticle dispersion, we measured its pH, which was $6.8 \pm 0.2$. For the 60 nm diameter particles and for a 1:150 diluted dispersion, we determined a zeta potential of $\zeta = -58 \, \text{mV}$, a specific conductivity of $\Lambda = 11.5 \, \mu\text{Scm}^{-1}$, and a hydrodynamic diameter of $2a = 62.1 \, \text{nm}$ using a Malvern Zetasizer. As expected for strong electrolytes, such as sodium citrate or sodium chloride, which are both present from the synthesis (*25*), we observed a linear dependence of the conductivity on the degree of dilution.

To estimate a Debye length from conductivity measurements of the dispersions, we assume that the citrate molecules are predominantly attached to the particles and the Debye length is governed by the cation concentration obtained from the conductivity measurement. The Debye length of all dispersions can be found in Table *S1*.

Sample preparation

For patterning, we used the thermally-sensitive resist polyphthalaldehyde (PPA). The material was synthesized by J. Hedrick and co-workers at IBM Research – Almaden in San Jose, CA, USA as described in (*33*).

Highly doped silicon wafers were used as substrates for PPA. To increase the adhesion of PPA in liquid environments, the wafers were first coated with a layer of HM8006 (JSR Inc.). The pre-formulated HM8006 solution was spin-coated at 3000 rpm and subsequently cured at 225 °C on a hotplate for 90 s for cross-linking. Then, the thickness of the HM8006 layer was determined by scratching the film with a scalpel and measuring the scratch profile by atomic force microscopy (AFM). From the profile, a thickness of approximately 52 nm was inferred. Next, PPA from a 6.5 weight percent solution in anisole was spin-coated on the sample. The coating process was adjusted such that a PPA layer thickness of 150 - 170 nm was achieved. After spin-coating, the wafer was baked at 90 °C for 2 min to remove any residual solvent. Note that the correct layer thickness is crucial for optimal particle contrast. The thicknesses of all samples used in our experiments can be found in Table *S1*.

Ratchet topography patterning

Two homemade setups were used for thermal scanning probe lithography (t-SPL) of the PPA films, see e.g. (*24*). The t-SPL tools comprise piezo scanners for fine-positioning of the tip with respect to the sample (Nano-MET2 (*x,y*) & Nano-OP30 (*z*), Mad City Labs Inc., USA, or P-733-2D (*x,y*) & P-753 (*z*), Physical Instruments). Only tips with a radius

of >5 nm were used for 3D patterning because they enable high thermal transport and minimal pickup of material. The tips were heated to ≈900 °C for patterning and pulled into contact from a height of ≈50 nm using a capacitive force pulse of 5 µs duration between the sample and the cantilever. During this time, the hot tip creates a well-defined void with a depth controlled by the applied voltage. The writing parameters were optimized during the raster scan of the programmed pattern by using the read-back information obtained in the retrace direction of the scan (*24*). This so-called closed-loop lithography enables uniform writing depths of ≈1 nm accuracy across the entire write field.

Nanofluidic confinement apparatus

The nanofluidic confinement apparatus (NCA) is described in detail in (*25*) and shown schematically in Fig. *S1A*. In brief, we use piezo-elements to position the silicon sample and the cover slip with nanometer precision. Picomotors allow us to parallelize the two surfaces with a precision of better than 1 nm across a lateral distance of 10 µm. The two surfaces can be approached with nanometer accuracy until single nanoparticles get into contact with both surfaces. For our experiments, a 20 - 30 µl droplet of nanoparticle dispersion is first placed onto the patterned PPA, and then the sample is inserted below the glass cover slip. The amount of liquid and the sample size of ≈1x1 cm$^2$ ensure that the droplet overflows the sample, which leads to a reduced curvature of the droplet (see $d_{men}$ in Fig. *S1A*). This reduces the capillary pressure in the slit, increasing the stability of the system. Water droplets are placed around the sample to locally increase the humidity and thus to minimize evaporation of the nanoparticle dispersion. As a result, we could perform experiments over more than 4 h without a noticeable change of the system parameters.

For good optical access to the region of interest and for an obstacle-free approach of both surfaces, the cover glass has a 40 - 45 µm high mesa at its center. To fabricate the mesa, a flat piece of glass (D263T borosilicate, UQG) was first cleaned in piranha solution for 5 min. Then, a protective layer of 30 nm Cr and 300 nm Au was sputtered on top. The layer was removed everywhere outside the intended mesa area by photo-lithography and wet etching (TechniEtch ACI2, MicroChemicals and TechniStrip Cr01, MicroChemicals). Subsequently, the glass was etched for 75 s by concentrated hydrofluoric acid (49% HF), resulting in a 40 - 45 µm high mesa. The residual metal layers were removed from the mesa using the same wet chemistry as before. Finally, the electrodes were added by vapor deposition of a 5 nm Cr adhesion layer and a 50 nm Au layer through a shadow-mask. The mask was made from a 200 µm steel plate by laser cutting (Lasercut AG).

Prior to use, the mesa was thoroughly cleaned using a peel-off polymer (Red First Contact, Photonic Cleaning Technologies) and either an oxygen or a hydrogen plasma at 200 W (GigaEtch 100-E, PVA TePla GmbH), and then rinsing it in deionized water. The process resulted in smooth and clean hydrophilic surfaces, as verified by AFM measurements and contact angle measurements, respectively.

Particle imaging and tracking

Imaging of particles was performed by interferometric scattering detection, (*25*, *27*), which provides an excellent signal-to-noise ratio for the detection of small particles. The details are described in (*25*). In brief, a collimated continuous-wave laser (50 mW, 532.1 ±0.3 nm, Samba, Cobolt) with a beam diameter of ≈0.7 mm was used as light source. The laser beam was focused onto the sample by a 100x, 1.4 NA oil-immersion objective

(Alpha Plan-Apochromat, Zeiss), resulting in a focal spot of ≈2 µm diameter. The focal spot was raster-scanned across the sample by a two-axis acousto-optic deflector (DTSXY, AA Opto-Electronic). The reflected light was collected by the same objective and guided onto a high-frame-rate camera (MV-D1024-160-CL-12, Photon Focus) using an 80:20 beam splitter.

To minimize the illumination time of the particles, the laser is typically raster-scanned only once along the *y*-axis for each frame. From the 1 µm spot size of the laser and the 500 nm spacing and 10 µs duration of the scanned laser lines along the *x*-axis, the time a particle interacts with the laser light is estimated to be about 40 µs. In this time interval, the particles move less than 10 nm at the highest observed drift speeds of 200 µs/s and diffuse on the order of 40 nm for a measured diffusivity of 4.6 µm$^2$s$^{-1}$. As both distances are small compared with the focal spot, we conclude that the image of the particle taken by the camera represents its position averaged over 40 µs.

Note that too long an illumination time of the camera can lead to incorrect results. However, as explained below, an illumination time of 40 µs does not lead to significant errors. To determine the lateral position of a particle, we calculated its radial symmetry center, which yields similar accuracies as Gaussian fitting (*25*). In our experiments, we typically observed a signal to noise ratio of ≈50:1, leading to a localization accuracy of better than 1.5 nm. The particle trajectories were reconstructed by linking the lateral particle positions of the individual frames (*25*).

Calculation of average drift velocities

To obtain average values of the drift velocities <*v*> in the electric field direction, we calculated the mean particle displacement Δ*x*$_i$ for time intervals Δ*t* ranging from 1 ms to the half period *τ*/2 of the electric field,

$$<\Delta x(\Delta t)> = \langle \frac{1}{N-1} \sum_{i=1}^{N-1} x(t_i + \Delta t) - x(t_i) \rangle = <v> \Delta t, \quad (S1)$$

where <...> signifies the ensemble average and *N* is the number of Δ*t* values per half cycle *τ*/2: N = *τ* / (2 Δ*t*). The value for <*v*> was then obtained by a linear fit of the data, see Fig. *S5*. The total drift velocities were obtained in the same manner, but for a time period of the full cycle time *τ* of the electric field.

FEM modelling: Physical model of the nanofluidic slit

Assuming thermal contact with a reservoir, the probability of observing a particle at a certain position *x*$_p$ is readily shown to be proportional to the corresponding free energy $F = U - TS$, with *U* being the internal energy and *S* the entropy. As the electrolyte concentration was less than 1 mM in our experiments, we only consider dilute solutions in the following. For such dilute solutions, a mean-field approach yields good agreement with more computationally intensive treatments based on Monte-Carlo methods and density functional theory. In the mean-field approximation, the most likely distribution of ions in an electrolyte satisfies the Poisson–Boltzmann equation. For a 1:1 electrolyte with a volume average concentration of each ion species of *n*$_0$ this is (*34*):

$$\nabla \cdot (\varepsilon_r \varepsilon_0 \nabla \phi) = 2 \, e \, n_0 \sinh \frac{e\phi}{k_B T} \,, \quad (S2)$$

4

where $\phi$ is the electrostatic potential, $\varepsilon_0$ is the permittivity of the vacuum, and $\varepsilon_r$ is the relative permittivity of the electrolyte.

If all charge in the domain that is not part of the electrolyte is assumed to be fixed, then the free energy for the system may be calculated from functionals on $\phi$ for the entropy of the electrolyte $S_S$ and the electrostatic energy $U_S$ of the charge configuration (*35*):

$$S_S = -2n_0 k_B \int \left( \frac{e\phi}{k_B T} \sinh \frac{e\phi}{k_B T} + 1 - \cosh \frac{e\phi}{k_B T} \right) dV \tag{S3}$$

$$U_S = \int \frac{\varepsilon_0 \varepsilon_r}{2} \nabla\phi \cdot \nabla\phi \, dV \tag{S4}$$

The model (see Fig. *S3*) was solved in three dimensions. The geometry and governing equation were symmetric with respect to the plane $y = 0$ so that only half of the domain was explicitly modelled. The symmetry boundary condition $\nabla\phi \cdot \vec{n} = 0$ was applied to the surface $y = 0$. Two periods of the ratchet were modelled, and a periodic boundary condition was applied to the pair of bounding faces of constant $x$. The boundary condition $\nabla\phi \cdot \vec{n} = 0$ was applied to the remaining surfaces. This enforced charge neutrality ($\int \rho \, dV = 0$) within the computational domain.

FEM modelling: Physical parameters used for modelling the nanoslit

The mean-field approach coupled with the assumption of constant charge means that all materials in the system are characterized by a relative permittivity, a surface charge, and, in the case of the electrolyte, its concentration $n_0$. The assumption of constant charge should be quite accurate. The thermal energy available to the particle ($1\,k_B T$) prevents it from approaching the ratchet surfaces too closely. For surface separations exceeding one Debye length, the effects of charge regulation can safely be neglected (*29*).

For the 60 nm-diameter gold particle, we determined the surface charge arising from the bound citrate anions using a measurement of the zeta potential of $-58\,\text{mV}$ (*25*). This measurement of the diffuse layer potential was converted into a surface charge density by iteratively solving the Poisson–Boltzmann equation in spherical coordinates until a surface charge was found that yielded the desired surface potential. The conducting nature of the gold sphere means that the electrostatic potential will be uniform over its volume. This was represented computationally by specifying a large relative permittivity. A value of 1000 was used in the calculation.

The dilute electrolyte was taken to have the relative permittivity of water at 21 °C (*36*). The electrolyte was composed of the strong 1:1 electrolyte sodium chloride and the strong 1:3 electrolyte sodium citrate. The relative concentrations of the components are difficult to establish experimentally. As we deal exclusively with negatively charged surfaces, which result in the accumulation of the monovalent $\text{Na}^+$ ions (and $\text{H}^+$), we neglect the fact that some of the co-ions have a valence of 3 rather than 1. We selected a concentration $n_0$ of the monovalent ions of 0.3 mM which corresponded to a Debye length $\kappa^{-1}$ of 17.5 nm ($\kappa^{-1} = \sqrt{\varepsilon_r \varepsilon_0 k_B T / 2 e n_0}$). This Debye length was found numerically to yield the dependence of effective potential amplitude on gap height observed in the experiment. The upper bounding surface of the ratchet was composed of silica, which has a relative permittivity of 3.9. To determine the surface charge for the silica surface, we used the

5

model outlined in (*34*). The pH of the electrolyte was measured to be 6.8 in the experiment. For a Debye length of 17.5 nm ($n_0 = 0.3$ mM), the self-consistent solution of the constitutive equation and the Grahame equation (equations 7 and 8 of (*34*)) yields a diffuse layer potential of –90 mV or equivalently a surface charge density of –5.9 mC/m².

There is a contribution to the system's free energy (via $U_s$) from the interior of the silica, which was included in our calculations. However, we did not include the small correction to the interior electric field obtained when accounting for the Stern capacitance of the silica surface.

The lower bounding surface of the ratchet was the polymer PPA. In the absence of specific data, we used a relative permittivity appropriate to polymers (PMMA) of 3.6. We did not find a value for the surface charge in the literature for PPA films immersed in water. We observed experimentally that particles are repelled by several nanometers from the mid-gap position towards the silica interface (*25*). Thus, we believe that the surface charge of the PPA film exceeds that of the silica film. However, in the absence of specific data, we simply selected the same surface charge (–5.9 mC/m²) for the PPA surface as that of the silica film. This should yield an underestimate of the ratchet potential. Note that the effective potential depends as (*37*) $e\phi_{\text{eff}}/k_B T = 4\tanh(e\phi_d/4k_B T)$ on the diffuse layer potential. As such, for a flat surface $\phi_{\text{eff}}$ saturates at $\approx 100$ mV. Hence our final result will not be too sensitive to the exact surface potential in this high potential regime. The geometry of the PPA was measured by AFM. It was approximated by the sawtooth profile shown in Fig. *S3*.

FEM modelling: Implementation

We performed all computations on a desktop computer using finite-element software. The calculation was performed using the Fenics (fenicsproject.org) software package. Meshes were prepared using Gmsh (gmsh.info), and solutions were inspected using Paraview (www.paraview.org).

The depth of the domain was taken to be the diameter of the particle plus $6\kappa^{-1}$ so that the particle's surface charge was effectively screened from this boundary. Second-order Lagrange elements were used with an element size of $0.3\kappa^{-1}$ for all points that were within $1.5\kappa^{-1}$ of a charged surface. The mesh was allowed to grow to $1.5\kappa^{-1}$ at points far from the charged surfaces (see Fig. *S4A*). A mesh-resolution study was performed to validate that this mesh density yielded an error in the relative calculated value of $\Delta F$ of less than $0.1\,k_B T$. The surface charge was represented using a weak contribution from the relevant interior facets. This approach is in contrast to applying a boundary condition on the fluid domain, which is only valid in situations where symmetry alone ensures that the external space is field-free.

FEM modelling: Results

Fig. *S4* shows the results of the numerical simulation. From the solution depicted in Fig. *S4A*, the free energy $F$ for a particular particle position ($x, z$) can be evaluated. Repeating this calculation over a grid of particle positions within the gap yields the free-energy landscape of the system. This is shown in Fig. *S4B*. For the ratchet system, only the value of the $x$ coordinate is of interest. The probability $P(x)$ that the particle occupies a given position $x$ can be evaluated by integrating over the probability distribution arising from the free energy shown in Fig. *S4B* with respect to *z*. The result of this integral is shown

in Fig. *S4C*. This relationship can be viewed as an effective potential $V_0(x)$ for the one-dimensional system and can be defined as $V_0(x) = -k_B T \log P(x)$. The result of this calculation is shown in Fig. *S4D*. Fig. *S4E* shows how these effective potentials can be scaled to obtain a single master curve. The dependence of the scaling factor on the gap height is shown in Fig. *S4E*. It is well fitted over this range of gap heights by the ansatz suggested by the linear superposition approximation: $A \exp(-d/2l)$. Our fitted value of $l$ of 18 nm is in close agreement with the Debye length of 17.5 nm used in the simulation and the one estimated in the experiments (≈12 nm, Table *S1*).

Characterization of the energy landscape in the sorting experiment

To obtain the energy landscape of the sorting device, the width and center positions of the deep ratchets were first determined in the camera coordinate system. Next, a histogram of the *y*-coordinates of all recorded particles inside the ratchets was made. It is energetically favorable for the particles to reside in the trenches leading to clearly visible maxima in the histogram at the position of the trenches, see Fig. *S9A*. The maxima were fitted by Gaussians from which both the center and the width (FWHM) of the respective deep ratchets were inferred. With this knowledge, it was possible to consider particles inside and outside the trenches separately. In Fig. *S9B*, histograms of the median relative brightness of particles inside and outside the trenches with trajectories extending over at least 20 frames are shown. For both cases, three different particle populations can be distinguished (indicated by the dashed lines). From SEM pictures, see Fig. *S10*, we knew that there were three different types of particles in our dispersion: spheres with 60 nm and 100 nm diameter and a small fraction of platelets with 100 nm diameter. Accordingly, we associated the brightest and most abundant particles with the 60 nm spheres, the medium bright and least abundant particles with the platelets, and the darkest particles with the 100 nm spheres. To confirm the validity of this assignment, we considered a dispersion containing only 100 nm nanoparticles in the NCA. We reduced the gap distance in the nanofluidic slit until the particles were hardly moving, and recorded the optical contrast in a movie. Next, we reduced the gap distance further until all particles that were under the pillar were pushed into the polymer and, by that, were immobilized. After a rinsing and drying step, we imaged the immobilized particles by SEM, see Fig. *S11*. The SEM pictures revealed that all particles have approximately the same size, but have different outlines. By comparing the two results, we could conclude that there is a strong correlation between the outline of the 100 nm particles and the observed relative brightness, see Fig. *S11*.

To determine the potentials experienced by the respective particles, we made histograms of the *x*-coordinates of their positions, see Fig. *S9C*. As in the main part of the paper, we considered average histograms using cross-correlation. For the 60 nm spheres, we made two histograms, one for the deep and one for the shallow ratchets. As the 100 nm particles practically do not jump between the deep and the shallow ratchets owing to a large energy barrier, we only made histograms for the trenches for these particles. From the histograms, the potential could simply be calculated using Boltzmann's theorem. Using the width of the deep ratchets, $w_t$, and of the shallow ratchets, $w_s$, the potential difference $\Delta V$ for the 60 nm particles as shown in Fig. *4A* was determined. Therefore, the ratio between the probability to find a 60 nm particle in the trench, $p_{in}$, and that to find it outside, $p_{out}$, had to be determined for some bin in the lowermost subfigure of Fig. *S9C*. From theses probabilities, the potential difference $\Delta V_{bin}$ at the position of the bin can be calculated as

$$\Delta V_{bin}/k_B T = \log\left(\frac{p_{in}}{p_{out}}\right) + \log\left(\frac{4w_s}{3w_t}\right). \tag{S5}$$

The potential difference $\Delta V$ is then given by the difference between the two minima of the potentials experienced by the 60 nm particles inside and outside the trenches.

Modelling of the sorting device

To assess the performance of the sorting device, we first establish an approximate model for the interaction energy seen by the particles in both ratchets. For the model we consider the electrostatic interaction between a sphere and a plane and use this energy to approximate the energy barriers involved in the ratchet geometry. The approximation is justified since the lateral extend of the shallow slope profile of ≈400 nm is much larger than the particle to ratchet distance (<80 nm). In order to obtain the final potential profile we use the experimentally determined potential (see Fig. *2E*) and scale it corresponding to the calculated energy barriers.

For our experiments, the electrostatic interaction of a sphere with a plane $W_{SP}$ is well described by the linearized Poisson-Boltzmann equation with constant potential on the surfaces. The interaction energy $W_{SP}$ is then given by (*25, 29*)

$$W_{SP}(z) = W_0 r e^{-z\kappa}, \tag{S6}$$

where $z$ is the distance of the particle surface to the plane and $\kappa^{-1}$ is the Debye length. $W_0$ is given by

$$W_0 = 4\pi\varepsilon\varepsilon_0 \psi_{P,eff} \psi_{S,eff}, \tag{S7}$$

where $\psi_{P,eff}$ and $\psi_{S,eff}$ are the effective surface potentials of the plane and the sphere, respectively.

We further assume that the surface potential of both confining walls (PPA (1) and glass (2)) of the nanofluidic slit is similar ($W_0 = W_1 = W_2$, see also section *FEM modelling: Physical parameters used for modelling the nanoslit*) and the particle therefore resides on average in the middle of the gap. Note, that in the linear approximation a difference in the potential is equivalent to a small shift of the plane's position.

Similar to the FEM based model, we determine the probability $P(x)$ to the find the particle at position $x$ by integrating the 2D probability density $p(x,z)$ along the $z$ coordinate:

$$P(x) = \int_r^{d(x)-r} p(x,z)\, dz = \int_r^{d(x)-r} C \cdot \exp\left(-\frac{(W_{SP}(z) + W_{SP}(d(x)-z))}{k_B T}\right) dz \tag{S8}$$

where $C$ is a normalization constant and $d(x)$ the gap distance at $x$.

The potential barrier $\Delta V$ caused by a ratchet tooth of height $h$ can then be approximated by equation *S8* as:

$$\Delta V = \ln\left(\frac{P(x|d(x)=d)}{P(x|d(x)=d-h)}\right). \tag{S9}$$

Using equation *S9*, the experimentally determined gap distance and the measured geometry of the ratchets (depth of the shallow ratchet 40 nm with tooth height 30 nm, and

8

depth of the deep ratchet 60 nm with tooth height 25 nm), we find good agreement with the experimentally measured potentials for $\kappa^{-1} = 16$ nm and $W_0 = 2.6 k_BT$ /nm. These values correspond to ratchet energies of 2 and $4.5 k_BT$ for the 60 nm particles in the deep and the shallow ratchet, $10 k_BT$ of the 100 nm particles in the deep ratchets, and potential barriers of $1.5 k_BT$ and $7 k_BT$ for the 60 nm and 100 nm particles to enter the shallow ratchet. The good agreement of these values with the experimentally measured values (see Fig. *4A*) justifies the use of this simple model to assess the performance of the sorting device.

Moreover, we assume that the force is directly proportional and the diffusivity indirectly proportional to the particle radius *r*, respectively. In reality the dependence of the force *F* on the particle radius will be more complex, however, only the dimensionless force $\tilde{F} = FL/\Delta V$ for a particular period *L* and potential barrier $\Delta V$ matters for the device. This dimensionless force can be easily adapted by changing the period *L* of the ratchet. Using the scaled ratchet potentials, equation *1* from the main manuscript, and the measured force at 4 V driving potential in the linear ratchets (Fig. *3B*), we calculate the average drift speed of a particle of radius *r* in the sorting device in dependence of the gap distance *d* and the relative width of the trenches $w_{r,t}$ (see previous section), see Fig. *S9A*. The plot clearly depicts the highly non-linear behavior of the sorting device. For the device geometry used in the experiments ($w_{r,t}=5.0\%$), the direction of the drift reverses at a critical radius $r_c \approx 41$ nm. This means that particles smaller than 41 nm radius are driven to the positive *x* direction and particles above 41 nm radius in the opposite direction. Optimal drift speeds would be expected at 35 and 50 nm radius, close to the experimentally used particle radii. Clearly, the steep slope of $\approx$-7 µm/s/nm at the zero crossing indicates the extraordinary sorting potential of the device.

Assessment of the sorting performance

The main benefit of sorting different particle populations into opposite directions is that the two populations are driven against opposite boundaries of the device. Therefore, the center of the device always marks the separation line of the two populations and the device footprint can be kept at a minimum. While the ratchet transports the particles by a distance proportional to time, diffusion scales with square root of time. Thus, as long as the device is sufficiently long and the drift speed is finite, at some point a steady state is reached with completely separated particle populations.

In the steady state, the diffusive particle current at each boundary $x_b$ back into the device area, $-D\partial_x \varrho(x_b)$, cancels the current of particles drifting into the reservoirs, $\varrho(x_b)v$. Thus,

$$\varrho(x_b)v = -D\partial_x \varrho(x_b), \tag{S10}$$

where $\varrho(x)$ is the particle density at position *x* and *v* the average drift speed of particles. As can be shown by direct calculation, the particle density decays exponentially towards the middle of the sorting device

$$\varrho(x) = \varrho_0 e^{-v|x_b-x|/D} \tag{S11}$$

with normalization constant $\rho_0$ and a characteristic decay length of $\lambda = D/v$. For the 60 nm gold spheres with a measured diffusion constant of $4.6$ µm$^2$/s and an average drift speed in

9

the sorting device of approximately 10 µm/s, a characteristic length of 0.5 µm is obtained. The same calculation for the 100 nm particles which had a drift speed of approximately 50 µm/s delivers a characteristic length below 0.1 µm.

For sorting two particle populations with a certain difference in size, the length of the sorting device must be large against the characteristic length λ for each population. Fig. *S13A* depicts λ for two particle populations with sizes of $r_c$ - 0.5 nm and $r_c$ + 0.5 nm as a function of the critical radius $r_c$ and the relative width of the deep ratchets. The critical radius $r_c$ is conveniently adjusted by an adaptation of the gap distance, as depicted in Fig. *S13B*.

Moreover, the rocking motion and the finite time of the particles spend in either ratchet will add to the diffusive boundary of the separated particle populations. Given the predicted energy barrier of $<4 k_B T$ for the particles with radii close to $r_c$ to transit into the shallow ratchet, we estimate the size of both effects to be on the order of 1 - 5 µm in the worst case.

As a result, we conclude that a perfectly fabricated 10 µm long device is capable to fully separate particle populations at 40.5 nm and 41.5 nm radius. Naturally, for much smaller particles and/or for a smaller radial difference, the characteristic length λ increases accordingly. Nevertheless, merely increasing the device dimensions to compensate for the increase in λ is expected to guarantee perfect sorting. Moreover, we found that the increase in λ at smaller dimensions is mainly due to a reduced dimensionless force $\tilde{F}$, which can be restored by simply increasing the period *L* of the ratchets as described above.

Clearly, for a real device, one has to consider the fabrication tolerances for the performance of the device. If the gap distance changes locally by Δ*d*, the zero crossing of the particle current shifts according to the relation shown in Fig. S*13B*. For t-SPL a fabrication precision of 1 nm has been demonstrated (*24*), which shifts the critical radius by ≈ 0.5 nm. Since the modelled device was by far not optimized in terms of geometry and applied force, we estimate that for an optimized device the fabrication precision will be the limiting factor, currently restricting the separation precision to approximately 1 nm radial difference.

**Supplementary Text**

Effect of particle-motion averaging caused by finite illumination time.

Averaging over the particle motion distorts the probability density from which the energy landscape is calculated. To quantify this effect, we modeled the motion of a Brownian particle in a sawtooth potential (left column of Fig. *S2A*) and in the experimentally determined potential (right column of Fig. *S2A*) and simulated different illumination times. By considering two slightly different potentials, we were able to ascertain that our results do not depend on the exact details of the shape of the potential. Following the experimental observations presented in the main part, the depth of the potentials was set to $5 k_B T$, where $k_B$ is Boltzmann's constant and *T* the temperature. Similarly, the periodicity was set to 550 nm, and the diffusion constant was chosen as $D = 4.3 \, \mu m^2/s$. The motion of 300 individual particles with randomly chosen initial positions was sampled in time steps of 100 ns and for a total propagation time of 5 s. For each trajectory, the average positions for time intervals of 4 µs, 40 µs, and 400 µs were calculated. From the probability density of the average positions, the respective potentials

10

were reconstructed and their minimum values were set to $0\,k_BT$. As Fig. *S2B* shows, the difference between the initial and the reconstructed potentials is negligible for an illumination time of $4\,\mu s$. The observed systematic error for $40\,\mu s$ is on the order of $0.1\,k_BT$, which corresponds to 2%. In comparison with the other experimental errors, this is still acceptable. If, however, the illumination time is increased to 400 µs, the errors are on the order of $0.5\,k_BT$ - $0.75\,k_BT$, which corresponds to 10 % - 15 %. Therefore, the systematic error that is caused by an illumination time of $40\,\mu s$ does not impose any limitation on the validity of the experimental results observed. However, the illumination time should not be increased.

Monte-Carlo Simulation implementation

A too long illumination time of the camera leads to a distortion of the experimental probability densities and hence to false values of the reconstructed potential. The size of this effect was quantified using simulated trajectories. Let us consider the equation of motion of a Brownian particle as given by the Langevin equation (*38*):

$$m\ddot{x} + \gamma\dot{x} = -V(x) + \xi(t), \tag{S12}$$

where *x* is the position of the particle, *m* its mass, *γ* the mobility, *V(x)* the external potential, and *ξ(t)* the thermal noise at time *t*. For particles of extremely small size in liquid environments, as in our case, the particle dynamics is normally well described by ignoring the inertia term (Smoluchowski approximation), which leads to

$$\gamma\dot{x} = -V(x) + \xi(t). \tag{S13}$$

The thermal fluctuations are modelled as Gaussian noise with vanishing mean obeying the fluctuation-dissipation relation:

$$<\xi(t), \xi(t')> = \frac{2k_BT}{\gamma}\delta(t-t'). \tag{S14}$$

To obtain a trajectory from equation *S14*, a random initial position $x_0$ in the interval [0, *L*] was chosen, and the Euler–Mayurama method for solving stochastic differential equations was used to determine the time-resolved particle positions for an evolution time of 5 s with 100 ns step size. Note that because of the stochastic character of equation *S12*, even if the initial value is the same, every trajectory will be unique, i.e., different from all other trajectories. Therefore, to estimate the size of the statistical scattering of the results, the procedure was repeated 300 times.

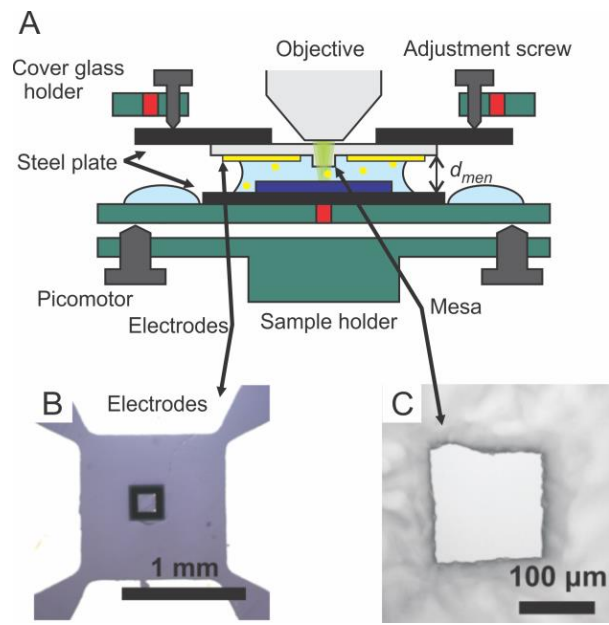

**Fig. S1**

**The nanofluidic confinement apparatus. A** Schematic of the mechanical setup. The sample (blue) and the cover glass are glued onto steel plates (black), which are magnetically attached to mechanical supports. The cover-glass holder allows alignment of the cover glass with respect to the focal plane of the objective. The cover-glass holder and the objective are mounted onto separate piezo fine positioners and a common coarse-positioning stage in vertical direction (not shown). During the experiments, the plane of the sample holder can be corrected using picomotors to enable parallelization of the two surfaces. The sample holder is attached to a coarse $(x,y)$-positioning system (not shown). **B** A square-shaped mesa of 30-50 µm height and 100-200 µm lateral dimension is located at the center of the cover glass. In the recessed area, Au electrodes are deposited to enable the application of electrical fields. **C** Optical image of the central mesa.

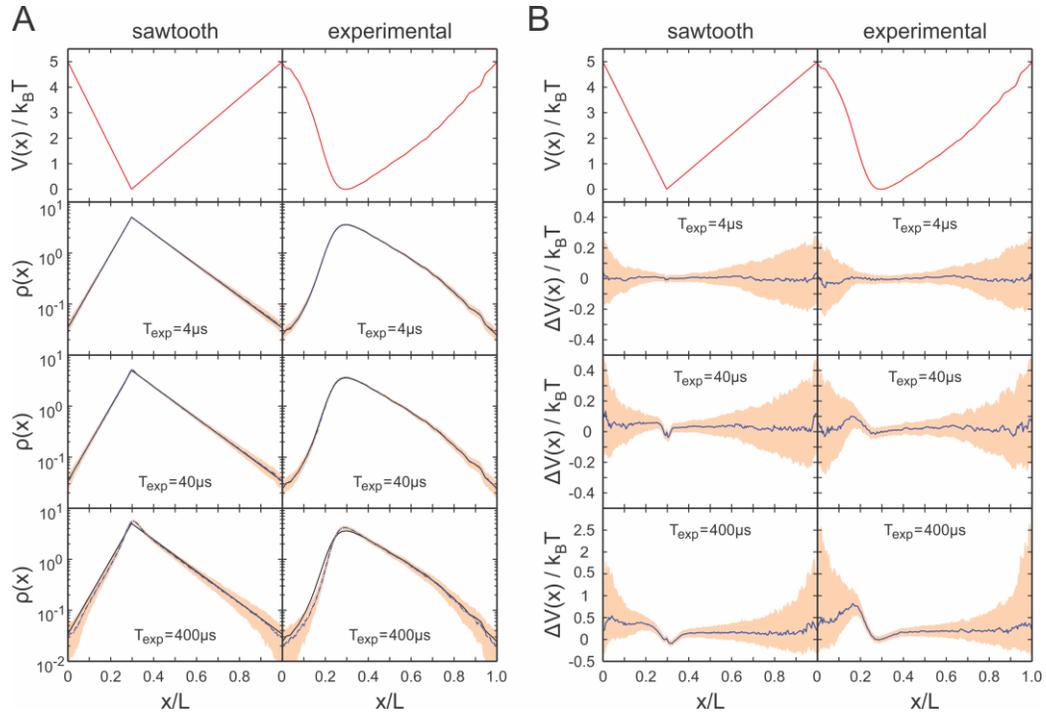

**Fig. S2**

**Monte-Carlo simulation results to investigate the influence of a finite averaging time on the reconstructed potentials**. **A** The periodicity of the potential $V(x)$ is set to $L = 550\,\text{nm}$, and the diffusion constant is chosen as $D = 4.3\,\mu\text{m}^2/\text{s}$. The scatter of the simulations is given by the shaded areas, which correspond to an uncertainty of one standard deviation. Initial potentials (top) and reconstructed probability densities $\rho(x)$ (bottom) for averaging times of 4, 40 and 400 µs. **B** Errors for the reconstructed potentials, same simulation parameters as in panel A. Note the different scaling of the y-axis for an illumination time of 400 µs.

Boundary Conditions

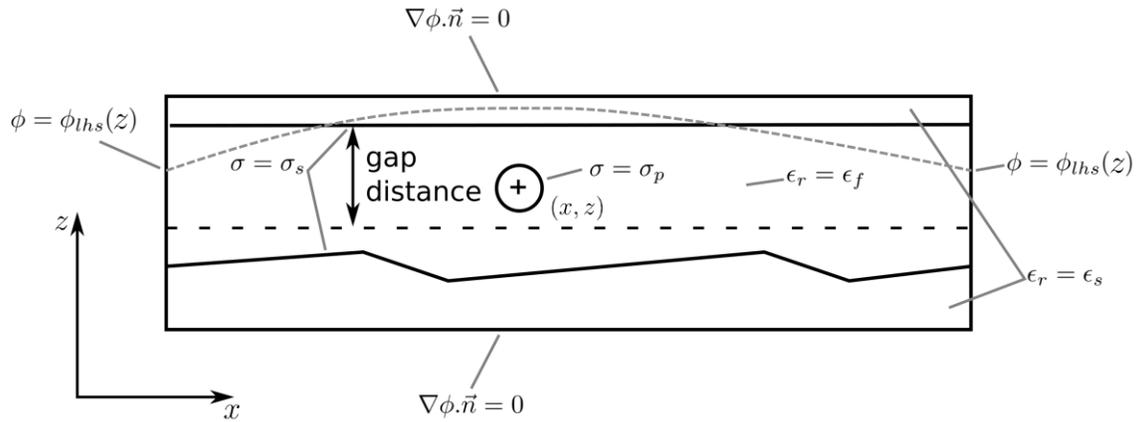

**Fig. S3**

**Modeling of the nanofluidic slit.** Geometry of the 3D model used to calculate the dependence of the effective potential seen by the gold particle on the gap distance. The boundary conditions and material parameters are shown by the labels.

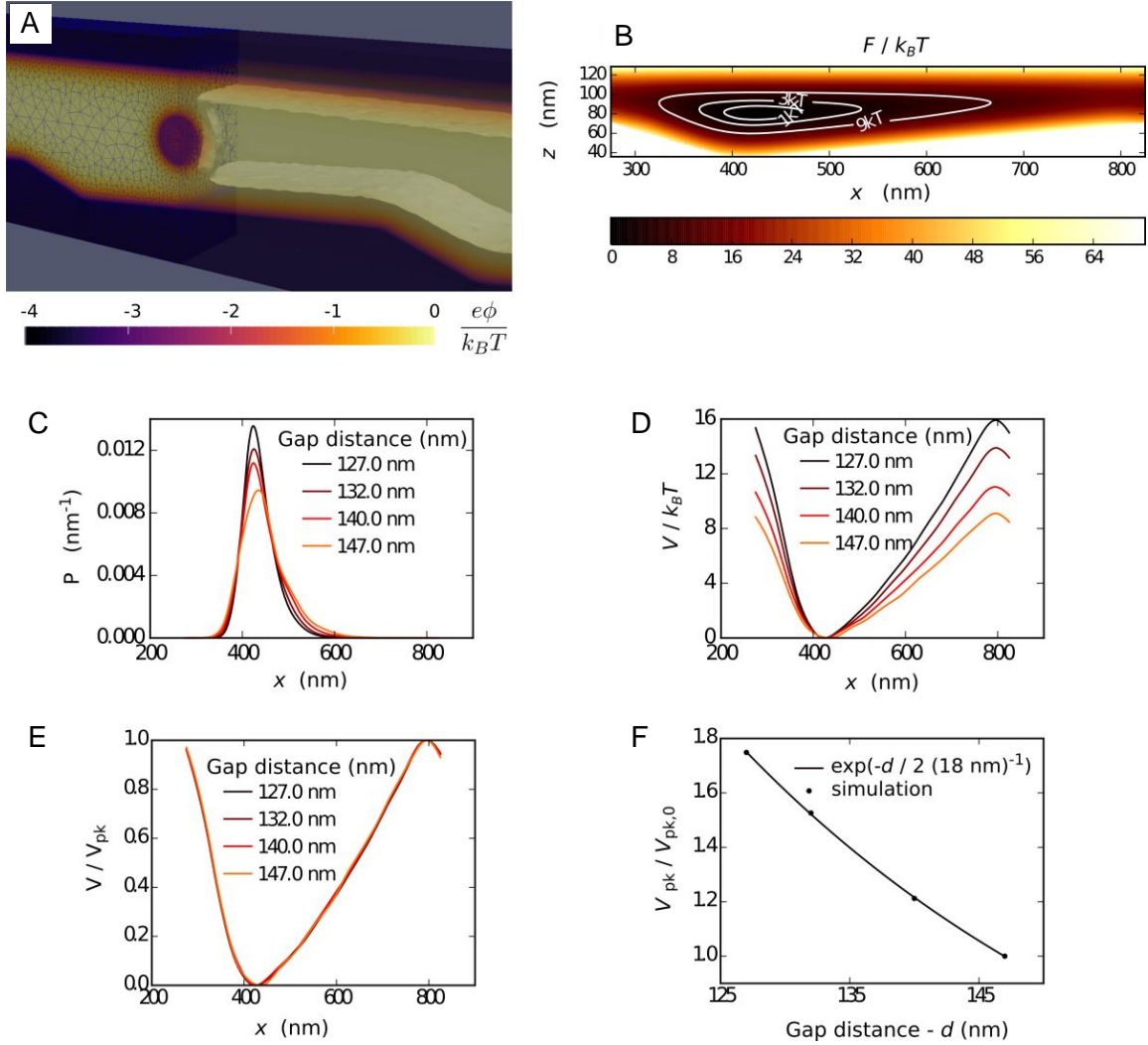

**Fig. S4**

**Results of the finite-element simulation of the effective one-dimensional potential experienced by the particle.** The gap distance *d* and particle position were varied, and all other independent variables were kept fixed. The simulation was performed for a Debye length $\kappa^{-1}$ of 17.5 nm (the full set of simulation parameters is given in the text). **A** The calculated electrostatic potential for a particle centered in the ratchet. The color scale encodes the electrostatic potential $e\phi/k_B T$. The white iso-surface corresponds to $e\phi/k_B T = -0.1$. The mesh used is shown in the left half of the plot. **B** The value of the free energy calculated for different particle positions $x_p$ in the ratchet for a gap height of *d* = 127 nm. **C** The calculated probability density on the particle position $x_p$. **D** The calculated effective potential corresponding to the probability density in panel C. **E** The result of re-scaling the curves in C by their maximum value. **F** The dependence of the maximum value of the effective potential on gap distance (dots). The solid line shows the result of a least-squares fit of the exponential function the simulated data. The fitted value of *l* was 18 nm.

15

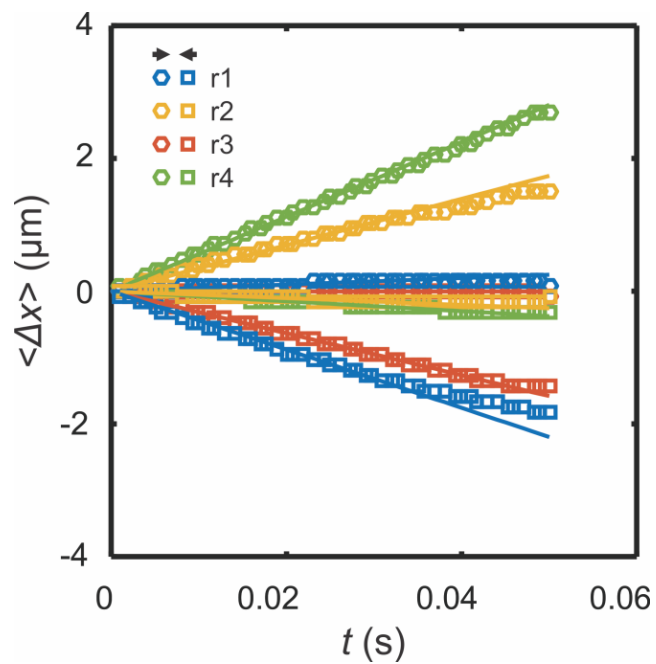

**Fig. S5**

**Measurement of the mean particle velocities.** The mean displacement $<\Delta x(\Delta t)>$ of the particles driven at 3.5 V @ 10 Hz at a gap distance of 127.5 nm along the *x*-direction, see Fig. 3A of the manuscript. Circles and squares indicate the two directions of the ratchets, respectively. The mean velocities are obtained from the linear fits shown by the solid lines.

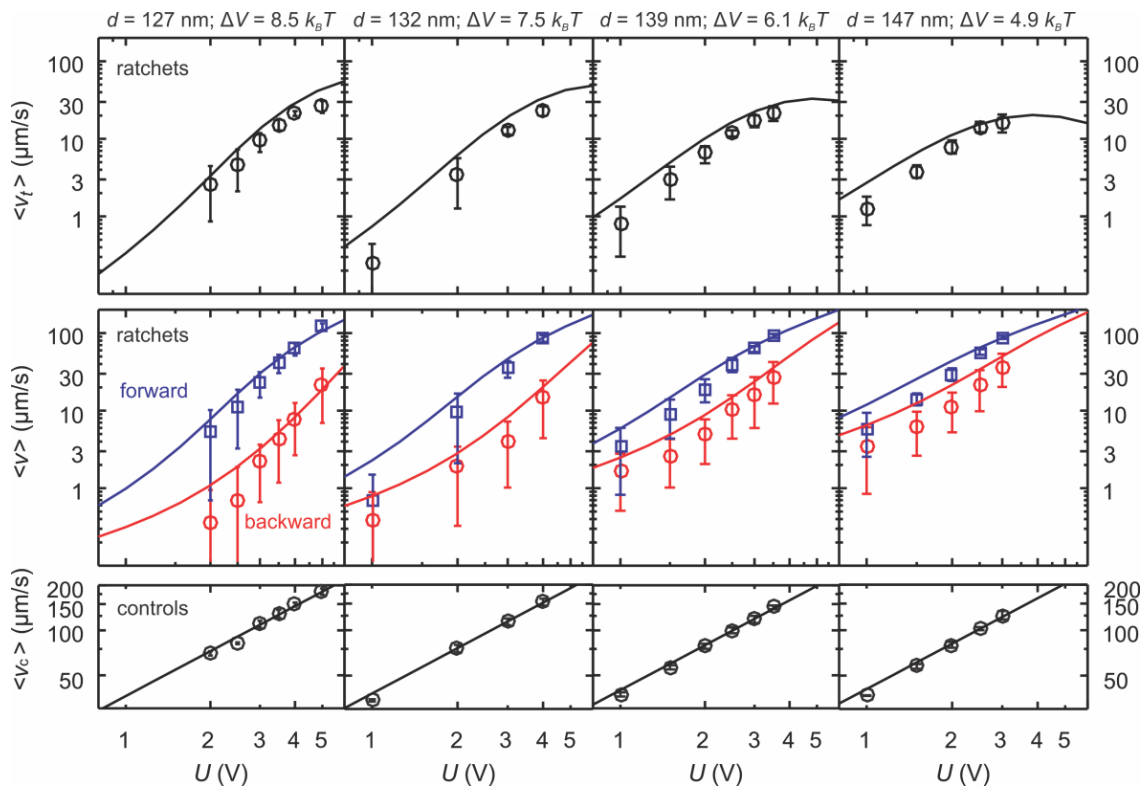

**Fig. S6**

**Average results for the particle drift as a function of applied voltage $U$ and for four gap distances.** From left to right, the gap distance $d$ is 127 nm, 132 nm, 139 nm, and 147 nm, corresponding to ratchet energy barriers of $\Delta V = 8.5\,k_BT$, $7.5\,k_BT$, $6.1\,k_BT$, and $4.9\,k_BT$. From top to bottom, the overall average drift speed $<v_t>$, the time-resolved average drift speed $<v>$ in the ratchets, and the average time-resolved drift speed $<v_c>$ in the control fields are shown. Lines are according to parameter-free model curves for the ratchets and best linear fits for the control fields.

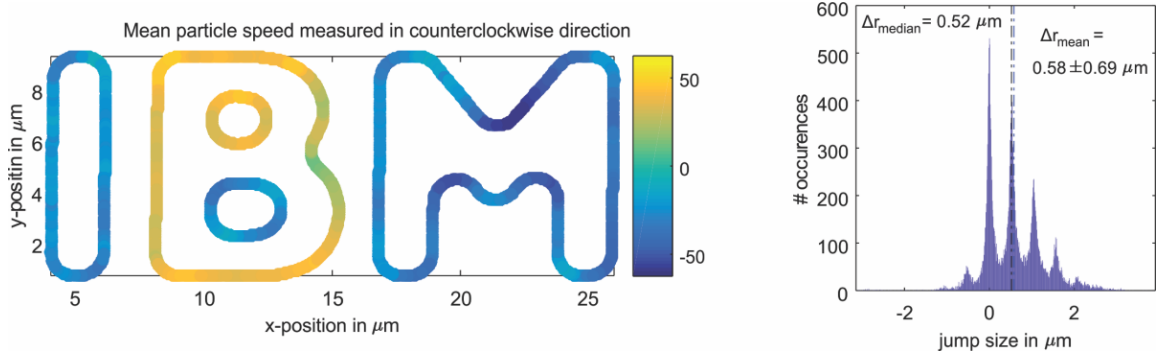

**Fig. S7**

**Measured particle speeds in curved paths.** **A** Measured average drift velocities for the ratchet topography shown in Fig. *1B*. To drive the particles, a square wave voltage of 4 V @ 30 Hz with a phase difference of 90° was applied to the two pairs of opposite electrodes located next to the mesa. **B** Histogram of the observed step-size distribution after each period of the applied field. A mean and a median step size of $0.58 \pm 0.69\,\mu$m and $0.53\,\mu$m were measured, respectively.

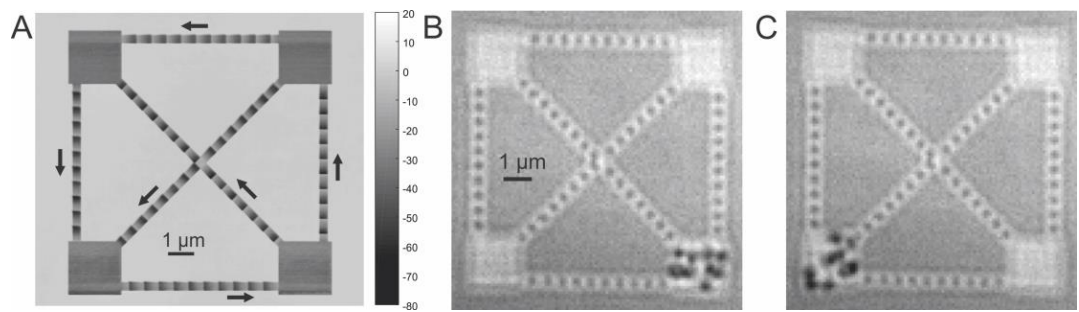

**Fig. S8**

**On-demand shuttling of 9 particles between reservoirs.** **A** Topography after t-SPL patterning. The ratchet directions are marked by arrows. The ratchet profiles ranged from 20 - 50 nm depth. **B** The 9 particles were transported into the lower right reservoir by a sequence of pulses. **C** Exploiting the directionality of the ratchets using aligned electric fields, all particles were routed to the lower left reservoir, see Movie S8.

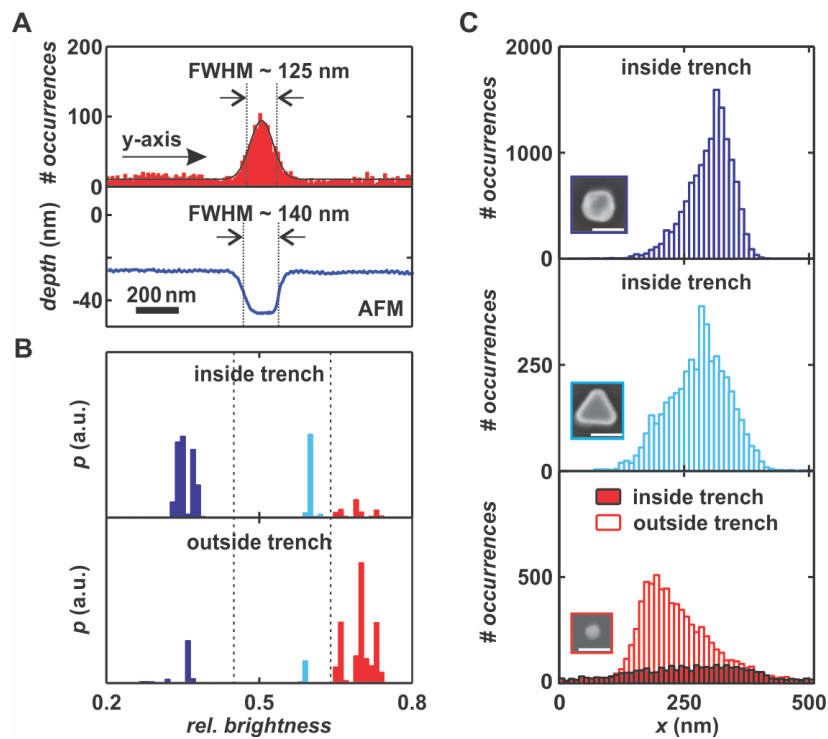

**Fig. S9**
**Characterization of the energy landscape of the sorting device. A** The width of the deep ratchets was determined to be 125 nm from a Gaussian fit of the particle-position histogram. The result is in good agreement with an average cross section measured by AFM shown in the lower part. **B** Three different particle populations could be distinguished by considering the median relative brightness of the particles. **C** For the different particle populations, average histograms were generated, from which the potential could be inferred using Boltzmann's theorem. Next to the histograms, SEM pictures of the particles that were associated with a certain brightness are shown. The size of the scalebars is 100 nm.

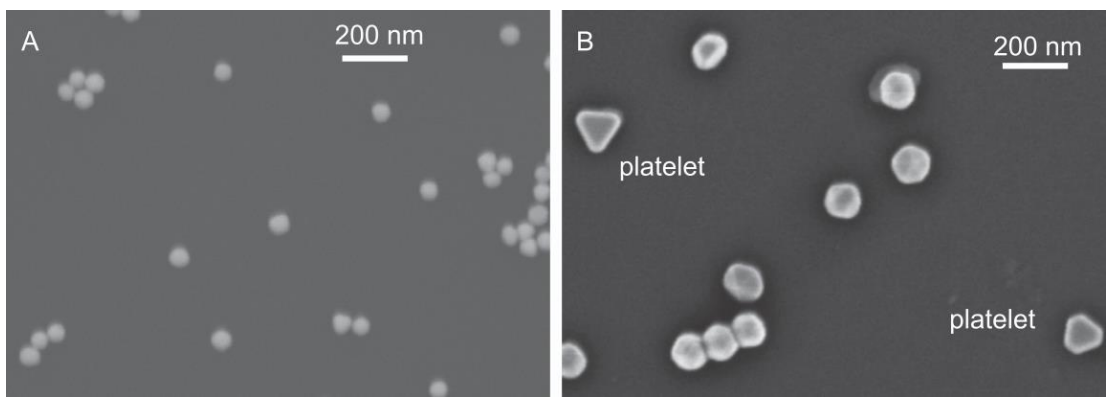

**Fig. S10**

**Particle populations. A** SEM image of randomly dispersed 60 nm-diameter Au nanospheres. The particle size distribution is almost monodisperse. **B** SEM image of randomly dispersed nominally 100 nm-diameter Au nanospheres. A significant fraction of triangular platelets is observed. Both images are at the same scale.

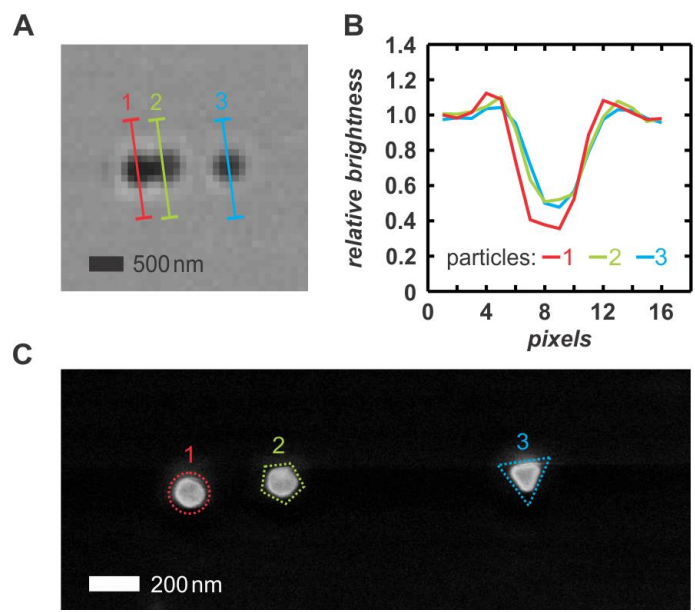

**Fig. S11**

**Particle outline and relative brightness.** **A** Three 100 nm nanoparticles as observed by the camera with the relative brightness measured along lines 1 to 3. **B** Particle 1 is considerably darker than particles 2 and 3, which have approximately the same relative brightness. **C** SEM image of the particles revealing their outline, which is spherical for particle 1, pentagonal for particle 2, and triangular for particle 3. Panels B and C corroborate the assignment between the relative brightness and the particle outline made in the main text.

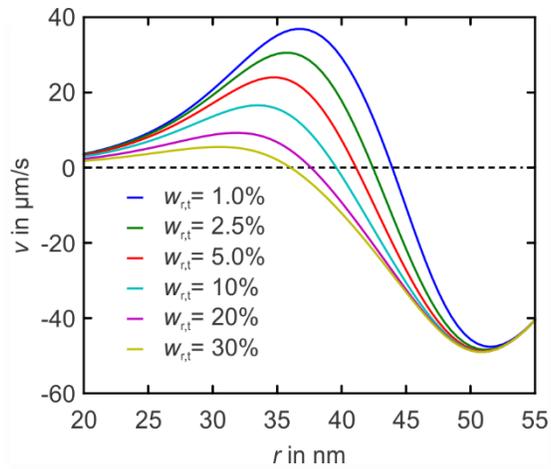

**Fig. S12**
**Drift speed vs particle radius.** Theoretically expected drift speeds for particles of different size and trenches of different relative width for a gap distance of $d = 150$ nm. Particles with radii smaller than the critical radius $r_c$ (given by the position of the zero crossing) are transported to the left, larger particles are transported to the right. In our experiment, the relative width of the trenches was approximately 5 % and thus the device was practically optimal for sorting $r_1 = 30$ nm and $r_2 = 50$ nm spheres as $r_c \approx 40$ nm.

23

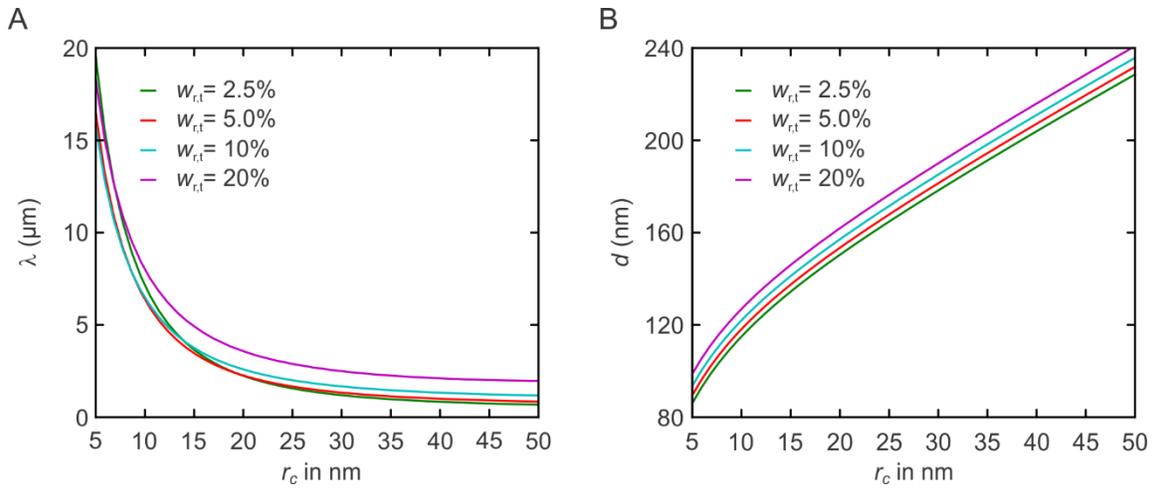

**Fig. S13**
**Particle sorting sensitivity. A** Simulated characteristic length λ for particles with radius $r = r_c \pm 0.5$ nm. The colored lines indicate four different relative widths of the trenches. **B** Gap distance required to adjust the critical particle radius to $r_c$ in panel A. Even for particles of only 5 nm radius, the gap distance is in the range between 80 nm and 100 nm which can easily be achieved by our experimental apparatus.

**Table S1**

Experimental parameters for the different Brownian motor experiments.

| Experiment | $t$ (HM) | $t$ (PPA) | Dilution | Est. $\kappa^{-1}$ | Gap distance $d$ |
|---|---|---|---|---|---|
| **Curved path** | $52 \pm 1$ nm | $160 \pm 2$ nm | 1:20 (60 nm) | 12.5 nm | $140 \pm 1$ nm |
| **Linear ratchets** | $52 \pm 1$ nm | $148 \pm 2$ nm | 1:18 (60 nm) | 11.9 nm | 127 - 148 nm |
| **On-demand transport** | $52 \pm 1$ nm | $160 \pm 2$ nm | 1:20 (60 nm) | 12.5 nm | $140 \pm 1$ nm |
| **Sorting** | $52 \pm 1$ nm | $160 \pm 2$ nm | 1:10 (60 nm) + 10:1 (100 nm) | | $150 \pm 0.5$ nm |

**Movie S1**

Real-time movie of Brownian motion of 6 diffusing 60 nm Au spheres in a ratchet geometry. The gap distance was $d = 147$ nm.

**Movie S2**

Real-time movie of rocking Brownian motor operation for 60 nm Au spheres at $d = 147$ nm and $U = 3$ V @ 10 Hz.

**Movie S3**

Real-time movie of rocking Brownian motor operation for 60 nm Au spheres at $d = 127$ nm and $U = 5$ V @ 10 Hz.

**Movie S4**

Slow-motion (factor 40) rocking Brownian motor operation for 60 nm Au spheres at 50 Hz ($d = 127$ nm and $U = 5$ V).

**Movie S5**

Slow-motion (factor 40) rocking Brownian motor operation for 60 nm Au spheres at 100 Hz ($d = 127$ nm and $U = 5$ V).

**Movie S6**

Slow-motion (factor 40) rocking Brownian motor operation for 60 nm Au spheres at 200 Hz ($d = 127$ nm and $U = 5$ V).

**Movie S7**

Slow-motion (factor 4.4) movie of rocking Brownian motor operation for 60 nm Au spheres along curved ratchet tracks. The electric field is rotating in counterclockwise direction in 90° steps starting at 45°.

**Movie S8**

On-demand shuttling of 60 nm Au particles between reservoirs. A sequence of electric field directions (bottom left) leads to the transfer of the particles into the bottom left reservoir.

**Movie S9**

Slow-motion (factor 3) movie for sorting of nominally 60 and 100 nm-diameter particles. An electric field in y-direction of 4 V @ 30 Hz is applied after ~5 s.



\end{document}